\documentclass{iopart}

\usepackage{cite}
\usepackage{epsfig}
\usepackage{iopams}
\usepackage{bm}
\usepackage{bbold}
\usepackage{color}
\usepackage{times}
\usepackage{braket}
\usepackage{cancel}
\usepackage[colorlinks,bookmarks=false,citecolor=blue,linkcolor=red,urlcolor=blue]{hyperref}
\hypersetup{%
  colorlinks = true,
  linkcolor  = black
}

\def\doi{http://dx.doi.org/}

\pdfoutput=1

\newcommand{\be}{\begin{equation}}
\newcommand{\ee}{\end{equation}}
\newcommand{\bea}{\begin{eqnarray}}
\newcommand{\eea}{\end{eqnarray}}
\newcommand{\nn}{\nonumber\\}
\newcommand\bes           {\begin{subequations}}
\newcommand\esu           {\end{subequations}}

\def\fr#1{(\ref{#1})}

\def\3pt#1#2#3{{\langle{#1}\vert{#2}\vert{#3}\rangle}}

\begin{document}

\title{On Truncated Generalized Gibbs Ensembles in the Ising Field Theory}
\author{F. H. L. Essler$^1$, G. Mussardo$^{2,3}$ and M. Panfil$^4$}
\address{$^1$ The Rudolf Peierls Centre for Theoretical Physics, Oxford University,
  Oxford OX1 3NP, UK}
\address{$^2$ SISSA and INFN, Sezione di Trieste, via Beirut 2/4, I-34151, 
Trieste, Italy}
\address{$^3$ International Centre for Theoretical Physics (ICTP),
I-34151, Trieste, Italy}
\address{$^4$ Institute of Theoretical Physics, University of Warsaw,
ul. Pasteura 5, 02-093 Warsaw, Poland
}
\begin{abstract}
We discuss the implementation of two different truncated Generalized
Gibbs Ensembles (GGE) describing the stationary state after a mass
quench process in the Ising Field Theory. One truncated GGE is based
on the semi-local charges of the model, the other on regularized
versions of its ultra-local charges. We test the efficiency of the two
different ensembles by comparing their predictions for the stationary
state values of the single-particle Green's function $G(x) = \langle
\psi^{\dagger}(x) \psi(0) \rangle$ of the complex fermion field
$\psi(x)$. We find that both truncated GGEs are able to recover
$G(x)$, but for a given number of charges the semi-local version
performs better.  
\end{abstract}
\maketitle

\frenchspacing

\section{Introduction}
\label{sec:intro}
It is by now well established that after a quantum quench integrable
lattice models relax locally to generalized Gibbs ensembles
(GGE)~\cite{RigolPRL07,CalabreseJStatMech07,CEF,FE,QAPRL,GGEgeneral},
\emph{cf.} Ref.~\cite{EF16} for a recent review.
The situation for integrable quantum field theories (QFT) is generally
more
complicated~\cite{CalabreseJStatMech07,qQFT,LL,pra,Cubero-Vernier}. A
key issue is that, in quantum field theories, expectation values of
ultra-local conservation laws typically suffer from ultraviolet
divergences and in order to define a GGE density matrix an appropriate
regularization procedure is needed. Here we follow the terminology of
Ref.~\cite{korepin} and call conservation laws ultra-local, if they
can be written as integrals of operators that act only at a point 
\be
Q_n=\int dx\ J_n(x)\ ,\quad n\in\mathbb{N}.
\ee
A related issue is that these ultra-local conservation laws are
generally not complete, in the sense that the GGE based on them may be
unable to describe the stationary state after an arbitrary global
quantum quench. In Ref.~\cite{pra} we have shown that in integrable
QFTs one can construct a different set of conserved charges that
fulfil a weaker form of locality. The densities ${\mathcal
  I}(x,x+\alpha)$  of these \emph{semi-local charges} \footnote{
In Ref.~\cite{pra} we called the conservation laws \fr{semilocal}
``quasi-local''. In order to avoid confusion with the use of this term
for lattice models, \emph{e.g.} see~\cite{Prosen:review}, we will
refer to them as semi-local in the following.  }
$I(\alpha)$ act non-trivially only on a given finite interval
$[x,x+\alpha]$  
\be
I(\alpha)=\int dx\ {\mathcal I}(x,x+\alpha)\ ,\quad \alpha\in\mathbb{R}^+.
\label{semilocal}
\ee
In contrast to the ultra-local conservation laws, the $I_\alpha$
typically do not to suffer from ultra-violet divergences. In the
integrable QFTs considered in \cite{pra} the completeness of the set
$\{I_\alpha\}$ derives from the fact that in the infinite volume the
charges are in one-to-one correspondence with the ``mode occupation
numbers'' $n(k)$ of the elementary excitations that exhibit purely
elastic scattering. The expectation values of the ultra-local charges
$Q_n$ encode only the moments of the mode distribution $n(k)$ and
therefore their knowledge may be not sufficient to reconstruct
$n(k)$. In this case the GGE built from the $Q_n$ would not be
complete. In integrable QFTs that support bound states (corresponding
to string solutions of the Bethe Ansatz equations) the situation is
more complicated and has been recently analyzed in
Ref.~\cite{Cubero-Vernier}.

In order to construct GGEs in practice, the concept of \emph{truncated
GGEs} introduced in Ref.~\cite{FE} has proven very useful. It is based on
approximating the full GGE by considering only certain subsets of
conserved charges. By choosing an appropriate sequence of such
subsets, the full GGE can then be defined by a limiting procedure. 
The main purpose of this note is to show how to implement a truncation
procedure for GGEs built from semi-local charges. To do this we focus
on the simplest possible case of a mass quench in the Ising field
theory. We complement these considerations by a construction of
truncated GGEs formed with regularized ultra-local charges. 

\section{Ising QFT and its conserved charges}

The Hamiltonian of the Ising field theory is
\be
H(m) = \int {\rm d}x\left(\frac{iv}{2}\left[R(x)\partial_x R(x) - L(x)\partial_x
    L(x)\right] + im R(x) L(x)\right)\ ,
\ee
where $R(x)$ and $L(x)$ are real right and left-moving fermion fields. The model is diagonalised by the Bogoliubov transformation 
\bea
  R(x) &= \int_{-\infty}^{\infty} \frac{dk}{2\pi} 
\sqrt{\frac{\omega(k)+vk}{2\omega(k)}}
  \left[e^{i\frac{\pi}{4}} Z(k)
    e^{-ixk} + {\rm h.c.} \right]\ ,\\
  L(x) &= \int_{-\infty}^{\infty} \frac{dk}{2\pi} 
\sqrt{\frac{\omega(k)-vk}{2\omega(k)}}
\left[e^{-i\frac{\pi}{4}} Z(k) e^{-ixk} + {\rm h.c.} \right]\,\,\,,
\label{Bogo}
\eea
where $\{Z(k),Z^\dagger(q)\}=2\pi \delta(k-q)$. The Hamiltonian takes the form
\be 
H(m) = \int_{-\infty}^\infty \frac{{\rm d}k}{2\pi} \underbrace{\sqrt{m^2 +
      v^2k^2}}_{\omega(k)} Z^{\dagger}(k) Z(k)\ .
\label{H}
\ee
It is obvious from \fr{H} that the mode occupation numbers
\be
N(k) = Z^{\dagger}(k)Z(k)
\ee
are conserved quantities. The Ising field theory has a set of conserved ultra-local charges of the form 
\bea 
\label{ultra-local}
  Q_n^{\pm} &=& \int_{-\infty}^{\infty} \frac{\rm dk}{2\pi} \epsilon_n^{\pm}(k) N(k),\\
  \epsilon_n^+(k) &=& \omega(k) k^{2n}, \quad\quad \epsilon_n^-(k) = v k^{2n+1}.
\eea
It follows from the expressions for $\epsilon_n^\pm(k)$ that expectation values of $Q^\pm_n$ for sufficiently large $n$ will generically suffer
from ultra-violet divergences. Moreover, the functions $k^m$ do not form a basis on the {\em infinite} interval $k \in (-\infty,+\infty)$,
which poses the question whether the set $\{Q^\pm_n\}$ is complete for constructing GGEs.

Applying the general construction of Ref.~\cite{pra} to the Ising field theory results in two continuous families of semi-local
conserved charges  
\bea \label{quasi-local}
  I^{\pm}(\alpha) &=& \int_{-\infty}^{\infty} \frac{dk}{2\pi} \epsilon^{\pm}(k,\alpha) N(k),\\
  \epsilon^+ (k,\alpha) &=& \omega(k) \cos(k\alpha) ,
  \quad\,\,\epsilon^-(k,\alpha) = \sin(k\alpha)\ .
\eea
These charges are clearly in one-to-one correspondence with the mode
occupation operators $N(k)$. It has been brought to our attention that
the fact that \fr{quasi-local} form a set of conserved charges for the
Ising field theory has been previously noted in Ref.~\cite{Doyonthesis}. 

\section{Mass quench}
In the following we focus on a particular quantum quench, the one in which we prepare the system in the ground state $|\Psi_0\rangle$ of
$H(m_0)$ and at time $t=0$ suddenly quench the mass parameter from $m_0$ to $m$. The time evolution following this quench is induced by
$H(m)$. This quench protocol has been previously studied in detail both on the lattice and in the continuum 
\cite{MajoranaFT,CEF,FE,QAPRL}. In order to work out the occupation number densities we put our system into a
large, finite box of length $L$. In this case, the finite volume Bogoliubov fermion creation/annihilation operators fulfil 
\be
\{Z^\dagger_L(k_n),Z_L(k_m)\}=L\delta_{n,m}\ ,\quad
k_n=\frac{2\pi n}{L}.
\ee 
Denoting the ground state of $H(m_0)$ in the finite volume by $|\Psi_0\rangle_L$, the mode occupation number densities are~\cite{MajoranaFT}
\be
\frac{1}{L}\frac{{}_L\langle
  \Psi_0|Z^\dagger_L(k_n)Z_L(k_n)|\Psi_0\rangle_L }
{{}_L\langle \Psi_0|\Psi_0\rangle_L}= \frac{|K(k_n)|^2}{1+|K(k_n)|^2}
\equiv n(k_n),
\label{N(k)_initial}
\ee
where
\be
  K(k) = \tan\left[\frac{1}{2}\arctan\left(vk/m\right) -
    \frac{1}{2}\arctan\left(vk/m_0\right)\right]\ .
\ee
We note that the mode occupation number densities are even functions of momentum, $n(k) = n(-k)$. In order to construct GGEs we require the
expectation values on the initial state $|\Psi_0\rangle_L$ of the conserved charges.  For the ultra-local conserved charges, the
expectation values suffer from ultraviolet divergences, which can be cured by regularizing the theory in terms of a sharp momentum cutoff
$\Lambda$. This means neglecting, in any physical quantity, all contributions coming from modes higher than $\Lambda$. As we are
dealing with a free theory, in which the modes are decoupled, this regularization does not spoil the integrability of the original
theory. The ``initial data'' is then given by 
\bea
i^\pm(\alpha)=\lim_{L\to\infty}\frac{1}{L}
\frac{{}_L\langle \Psi_0|I^\pm(\alpha)|\Psi_0\rangle_L}
{{}_L\langle \Psi_0|\Psi_0\rangle_L}&=&
\int_{-\infty}^{\infty} 
\frac{dk}{2\pi} \epsilon^{\pm}(k,\alpha)\ n(k),\nn
q^\pm_{n,\Lambda}=\lim_{L\to\infty}\frac{1}{L}
\frac{{}_L\langle \Psi_0|Q^\pm_{n,\Lambda}|\Psi_0\rangle_L}
{{}_L\langle \Psi_0|\Psi_0\rangle_L}&=&
\int_{-\Lambda}^{\Lambda} \frac{dk}{2\pi} \epsilon^{\pm}_n(k)\ n(k).
\label{initialdata}
\eea
Here we have defined
\be
Q^\pm_{n,\Lambda}\,=\,\int_{-\Lambda}^{\Lambda} \frac{\rm dk}{2\pi}
\epsilon_n^{\pm}(k) N(k) \ .
\label{ULreg}
\ee
We note that, unlike the original ultra-local charges $Q^\pm_n$, their regularized versions are no longer ultra-local but have instead densities that act non-trivially on the entire system. Moreover, in the presence of the cut-off, the functions $\epsilon^\pm_n(k)$ form a basis on the finite interval $[-\Lambda,\Lambda]$, and therefore for the cutoff theory we expect that the charges $Q^\pm_{n,\Lambda}$  are also able to construct a GGE. It is then interesting to compare the results obtained from the GGE's constructed from the two sets of conserved charges, namely the semi-local ones and the regularized version of the ultra-local charges. 

As $n(k)$ is an even function the expectation values of $I^-(\alpha)$ and $Q^-_{n,\Lambda}$ are identically zero, and these charges do not play a role in the respective GGEs, \emph{cf.} Ref.~\cite{FE}. From here on we focus on the following Green's function (other Green's functions can be analysed in the same way)  
\be\fl
G(x,t) = \lim_{L\to\infty} 
\frac{{}_L\langle\Psi_0(t)|\psi^{\dagger}(x)\psi(0)|\Psi_0(t)\rangle_L}{{}_L\langle\Psi_0|\Psi_0\rangle_L}, \qquad \psi(x) = \frac{1}{\sqrt{2}}(R(x) + i L(x)).
\ee
At late times, $G(x>0,t)$ approaches stationary values
\be \label{g(x)_through_N(k)}
\frak{g}(x) = \lim_{t\to\infty}G(x,t)=
\int_0^\infty \frac{dk}{2\pi}
\frac{m\cos(kx)}{\omega(k)}\big(2n(k)-1\big)\ ,\quad x>0.
\ee
The question we are interested is how well $\frak{g}(x)$ can be
described by the two different kinds of truncated GGEs based on the
sets $\{I_\Lambda(\alpha)\}$ and $\{Q_{n,\Lambda}\}$ respectively.

\subsection{Truncated GGE built from semi-local charges}
\begin{figure}[ht]
\begin{center}
\includegraphics[width= 0.5\columnwidth]{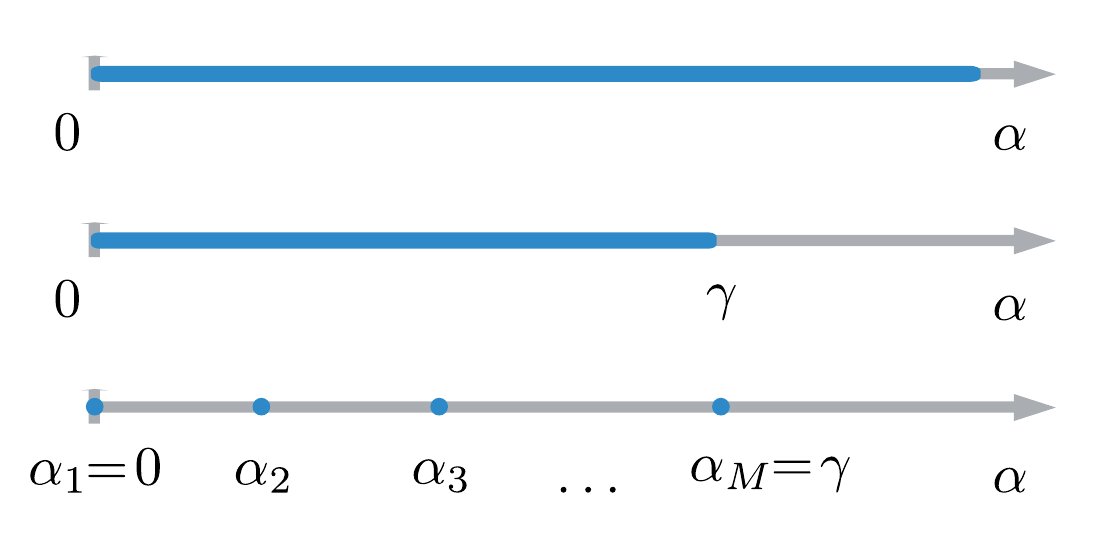}
\end{center}
\caption{The truncation scheme of the semi-local GGE is implemented
  firstly imposing a fixed cut-off in $\alpha$, denoted by $\gamma$,
  and then discretising the interval $[0,\gamma$] in $M$ equidistant
  points.}  
\label{fig:discetization}
\end{figure}

 Following the rationale of Ref.~\cite{FE}, we now wish to construct a
 truncated GGE by retaining only a finite subset of the conserved
 charges $I^+(\alpha)$ 
\bea
\rho_{{\rm sl},M} &=& \frac{1}{\mathcal{Z}_{{\rm sl},M}} \exp \left( - \sum_{m=1}^M
\lambda^+(\alpha_m) I^+(\alpha_m) \right)\ ,
\label{trhosl}
\eea
where the Lagrange multipliers are fixed by the conditions
\be
\fl
\lim_{L\to\infty}
\frac{1}{L}\frac{{}_L\langle \Psi_0| I^{+}(\alpha_m)|\Psi_0\rangle_L}
{{}_L\langle\Psi_0|\Psi_0\rangle_L} = \lim_{L\to\infty} \frac{1}{L}
{\rm tr}\left(\rho_{{\rm sl},M} I^{+}(\alpha_m)\right),\quad
m=1,2,\dots M.
\label{truncated_GGE_QL}
\ee
Eqns \fr{truncated_GGE_QL} do not require regularization except for
$\alpha_m=0$, which corresponds to the energy density and which is
ultraviolet divergent for the mass quench considered. It is important
to take the energy density into account, as the semi-local
conservation laws with $\alpha\neq 0$ cannot distinguish between
eigenstates characterized by a density $n(k)$ and eigenstates with 
density $n(k)+c/\omega(k)$ because
\be
\int_{-\infty}^{\infty} \frac{{\rm d}k}{2\pi} \epsilon^+(k, \alpha) 
\frac{c}{\omega(k)}=0\ ,\quad \alpha\neq 0.
\label{ambiguity}
\ee
Given that the energy density is ultraviolet divergent for the mass
quench, one could simply replace it by the density itself. This would
remove the ambiguity \fr{ambiguity}. However, in order to solve the
initial conditions \fr{truncated_GGE_QL}, it is in practice very
helpful to introduce a cutoff. We therefore proceed by introducing
semi-local charges with a cutoff $\Lambda$
\bea
I^\pm_{\Lambda}(\alpha)\,&=&\,\int_{-\Lambda}^{\Lambda} \frac{\rm dk}{2\pi}
\epsilon^{\pm}(k, \alpha) N(k),
\label{SLreg}
\eea
and then solve the initial conditions
\be
\fl
\lim_{L\to\infty}
\frac{1}{L}\frac{{}_L\langle \Psi_0| I^{+}_\Lambda(\alpha_m)|\Psi_0\rangle_L}
{{}_L\langle\Psi_0|\Psi_0\rangle_L} = \lim_{L\to\infty} \frac{1}{L}
{\rm tr}\left(\rho_{{\rm sl},M} I^{+}_\Lambda(\alpha_m)\right),\quad
m=1,2,\dots M.
\label{truncated_GGE_QL_cutoff}
\ee
This set of coupled nonlinear equations can be efficiently solved
numerically by increasing the number of charges $M$ by one at a time.
The set $\{\alpha_k|k=1,\dots,M\}$ that defines the truncated GGE is
conveniently chosen as shown in Fig.~\ref{fig:discetization}  
\be
\alpha_j=(j-1)\frac{\gamma}{M-1}\ ,\quad j=1,\dots,M.
\ee

Once the Lagrange multipliers $\lambda+(\alpha_m)$ have been computed
we can evaluate the expectation value 
\bea
\fl
\frak{g}_{{\rm sl},M}(x) &= {\rm tr}\Big(
\rho_{{\rm sl},M}\psi^{\dagger}(x)\psi(0)\Big)
=\int_0^\infty \frac{dk}{2\pi}
\frac{m\cos(kx)}{\omega(k)}\big(2n_{{\rm sl},M}(k)-1\big)\ ,
\quad x>0,\nn
\fl
n_{{\rm sl},M}(k) &= \left[
  1+\exp\left(-\sum_{m=1}^M\lambda^+(\alpha_m)\epsilon^+(k,\alpha_m)\right)\right]^{-1}.
\label{gm}
\eea

\subsection{Truncated GGE built from (regularized) ultra-local charges}
One can repeat the construction of the previous section but this time using the regularized ultra-local charges~\fr{ULreg}. The density matrix is
now  
\be
  \rho_{{\rm ul},M} = \frac{1}{\mathcal{Z}_{{\rm ul},M}}\exp\left(-\sum_{m=1}^M
  \lambda_m^+ Q_{m,\Lambda}^+\right), 
  \label{pippo}
\ee
and the Lagrange multipliers $\lambda_m^+$ are fixed by the conditions 
\be
\fl
\lim_{L\to\infty}
\frac{1}{L}\frac{{}_L\langle \Psi_0| Q^{+}_{m,\Lambda}|\Psi_0\rangle_L}
{{}_L\langle\Psi_0|\Psi_0\rangle_L} = \lim_{L\to\infty} \frac{1}{L}
{\rm tr}\left(\rho_{{\rm ul},M} Q^{+}_{m,\Lambda}\right),\quad
m=1,2,\dots M.
\label{truncated_GGE_UL}
\ee
A direct numerical solution of \fr{truncated_GGE_UL} is difficult,
because the expectation values of $Q_{m,\Lambda}^+$ increase rapidly
with the index $m$ and there is a lack of continuity in the values of
the Lagrange multipliers $\lambda_m^+$ with respect to the total
number of charges, i.e. the solutions $\{\lambda^+_m\}$ for the ensembles with
$M$ and $M+1$ charges will differ significantly. In order to obtain a
stable procedure we find it convenient to work with particular linear
combinations of the charges $Q^+_{n,\Lambda}$  
\be
\widetilde{Q}_m^+ = 2\int_0^{\Lambda} \frac{dk}{2\pi} \omega(k)\ 
T_{2(m-1)}(k/\Lambda) N(k)\ ,
\ee
where $T_m(x)$ are Chebyshev polynomials. In practice we re-express
\fr{pippo} and \fr{truncated_GGE_UL} in terms of these linear
combinations and solve for the corresponding Lagrange
multipliers. From these we can then determine the $\lambda^+_m$. The
Green's function takes the same form as in~\fr{gm} but now with the
ultra-local density matrix  
\bea
\fl
\frak{g}_{{\rm ul},M}(x) &= {\rm tr}\Big(
\rho_{{\rm ul},M}\psi^{\dagger}(x)\psi(0)\Big)
=\int_0^\Lambda \frac{dk}{2\pi}
\frac{m\cos(kx)}{\omega(k)}\big(2n_{{\rm ul},M}(k)-1\big)\ ,\quad x>0,\nn
\fl
n_{{\rm ul},M}(k) &= \left[
  1+\exp\left(-\sum_{m=1}^M\lambda_m^+\epsilon_m^+(k)\right)\right]^{-1}.
\label{gul}
\eea

\section{Results} 
In the following we present results for two quench processes 
(i) a \emph{small quench} $m_0=m/2\longrightarrow m$;
(ii) a \emph{large quench} $m_0=5m\longrightarrow m$.
Here large/small is measured by the density of excitations $n(k)$
after the quench, \emph{cf.} Refs~\cite{CEF}. In order to better
distinguish between the different ensembles we focus on the
dimensionless mode-distribution dependent part of the Green's
functions 
\be
g(x) = 2v\int_0^{\infty} \frac{{\rm d}k}{2\pi}
\frac{\cos(kx)}{\omega(k)}n(k). 
\ee
The cutoff $\Lambda$ entering our construction of the ensembles is chosen such that in the spatial range considered in our plots the
function 
\be
g_\Lambda(x) = 2v\int_0^{\Lambda} \frac{{\rm d}k}{2\pi}
\frac{\cos(kx)}{\omega(k)}n(k)
\ee
reproduces the full answer $g(x)$ up to errors that are negligible on the scale of our plots. The values we use are $\Lambda = 5m/v$ for the
small quench and $\Lambda=20m/v$ for the large quench.

\subsection{Truncated ultra-local GGE}
Our truncated ultra-local GGEs are characterized by the number $M$ of conserved charges and the momentum cutoff $\Lambda$. We first analyze
the dependence on $M$ for a fixed value of $\Lambda$. 
\begin{figure}[ht]
  \begin{center}
     \includegraphics[width= 0.65\columnwidth]{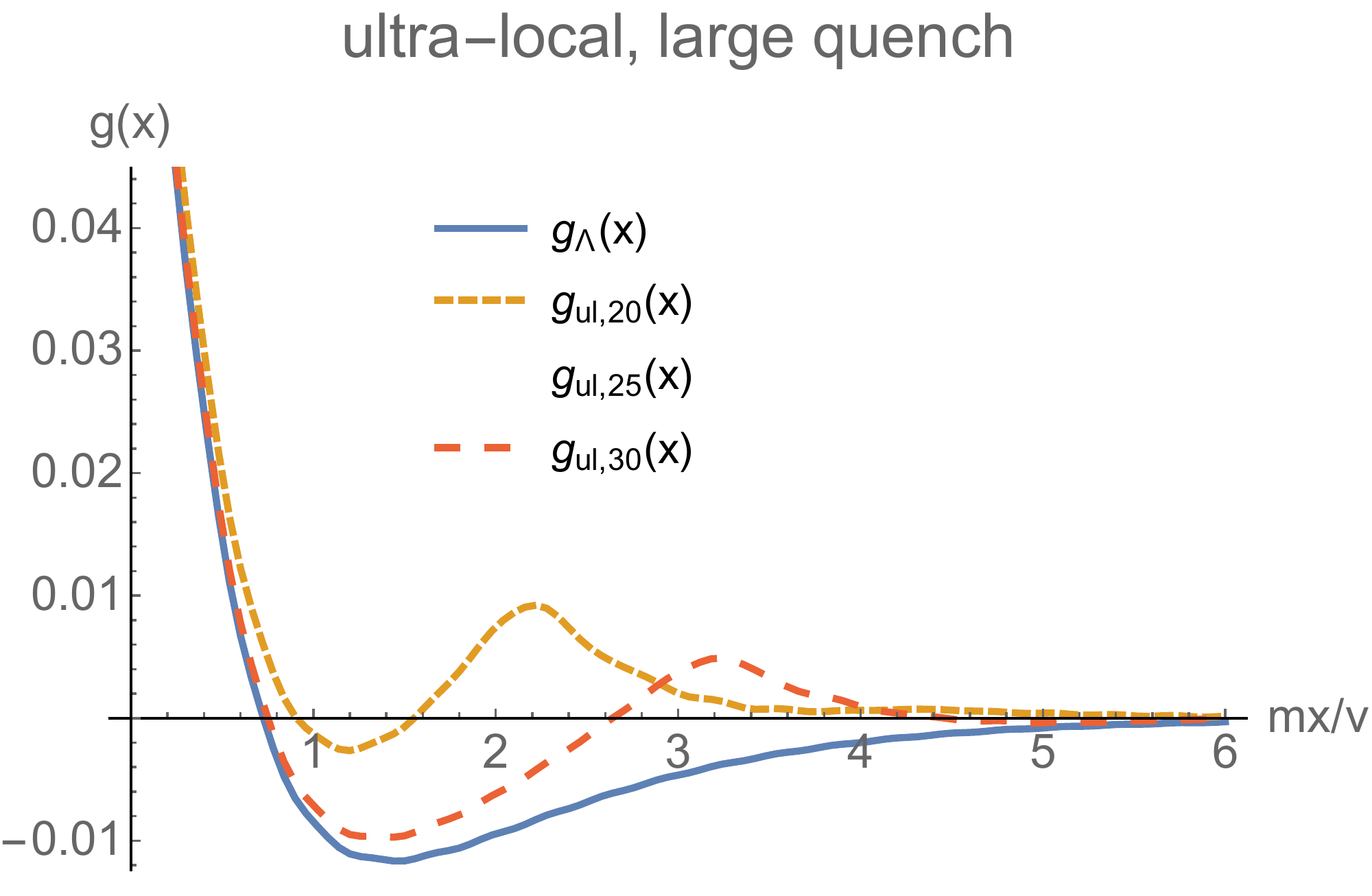}
  \end{center}
\caption{Truncated ultra-local GGE predictions $g_{{\rm ul},M}(x)$ for increasing numbers $M$ of ultra-local charges. Here the cutoff is
fixed at $\Lambda=5m/v$. We see that the range in $mx/v$ over which the truncated GGE provides an accurate description of the stationary state
result $g_\Lambda(x)$ grows as $M$ is increased. 
} 
\label{fig:ultra_local_M}
\end{figure}

\begin{figure}[ht]
  \begin{center}
    \includegraphics[width= 0.65\columnwidth]{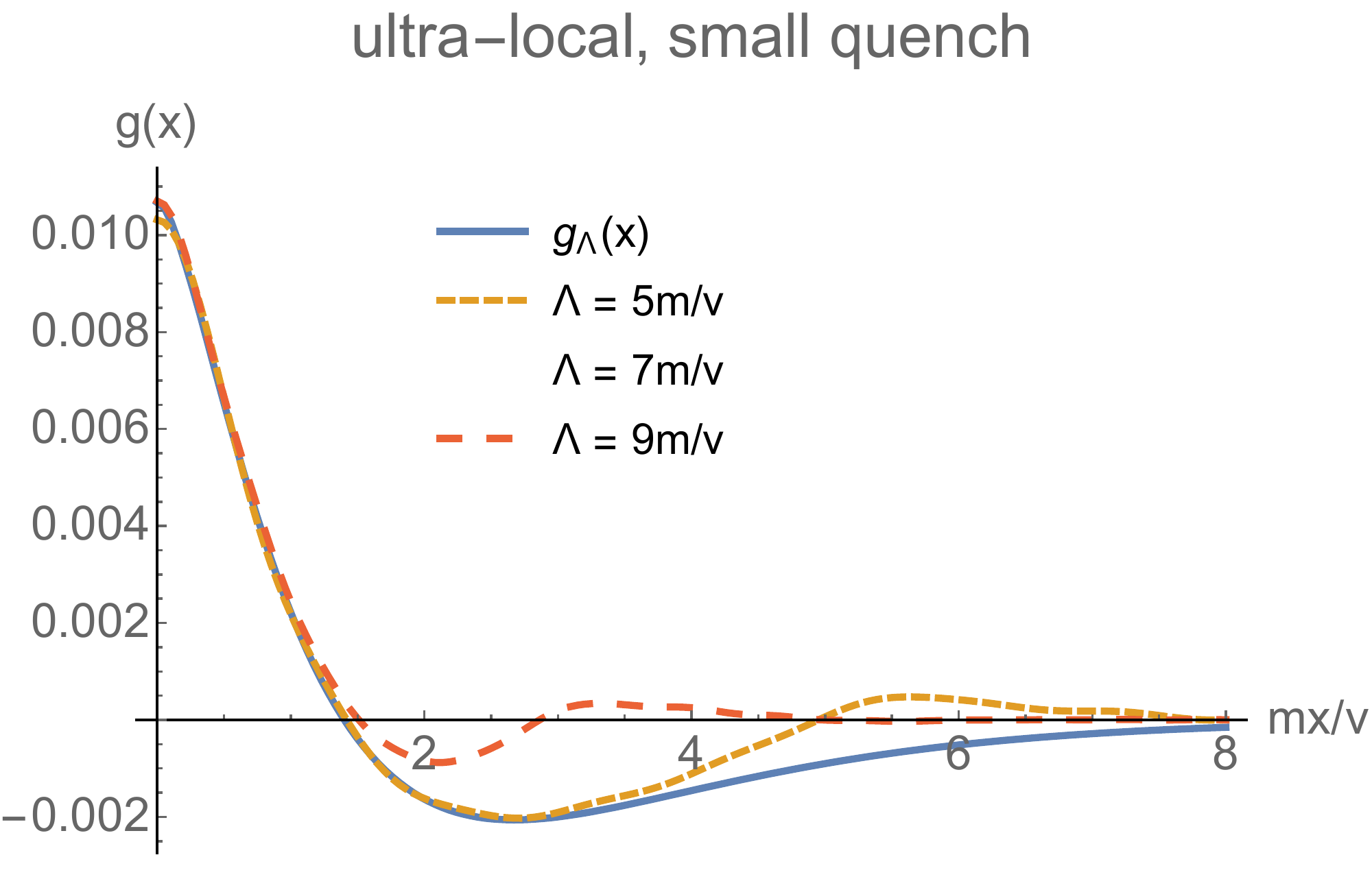}
  \end{center}
\caption{Dependence of the truncated ultra-local GGE predictions $g_{{\rm ul},M}(x)$ with $M=15$ charges on the cutoff $\Lambda$. 
} 
\label{fig:ultra_local_Lambda}
\end{figure}

Fig.~\ref{fig:ultra_local_M} shows the effect of increasing the number of charges $M$ employed in the ultra-local GGE, while keeping the
cutoff fixed. We see that increasing $M$ extends the region in $mx/v$, over which the truncated GGE provides an accurate approximation 
$g_{\rm UL,M}(x)$ of the stationary Green's function $g_\Lambda(x)$.
We next investigate the cutoff dependence at a fixed number $M$ of
charges. We consider a small quench so that a small number of charges
already provides good results. As shown in
Fig.~\ref{fig:ultra_local_Lambda} increasing $\Lambda$ at fixed $M$
(and of course adjusting the Lagrange multipliers accordingly) gives a
truncated GGE that provides a good description of $g(x)$ over a
\emph{decreasing} range in $mx/v$.

We may summarize our results for truncated ultra-local GGEs as
follows:
\begin{enumerate}
\item{} In order to obtain an accurate description of $g(x)$ on a given range $[0,x_0]$ we require a sufficiently large cutoff
$\Lambda$. 
\item{} Given $\Lambda$ one needs to increase the number of ultra-local charges $M$ until convergence of $\frak{g}_{{\rm ul},M}(x)$ 
on the interval $[0,x_0]$ has been achieved. The larger $\Lambda$, the higher $M$ needs to be.
\end{enumerate}

\subsection{Truncated semi-local GGE}
\begin{figure}[ht]
  \begin{center}
    \includegraphics[width= 0.65\columnwidth]{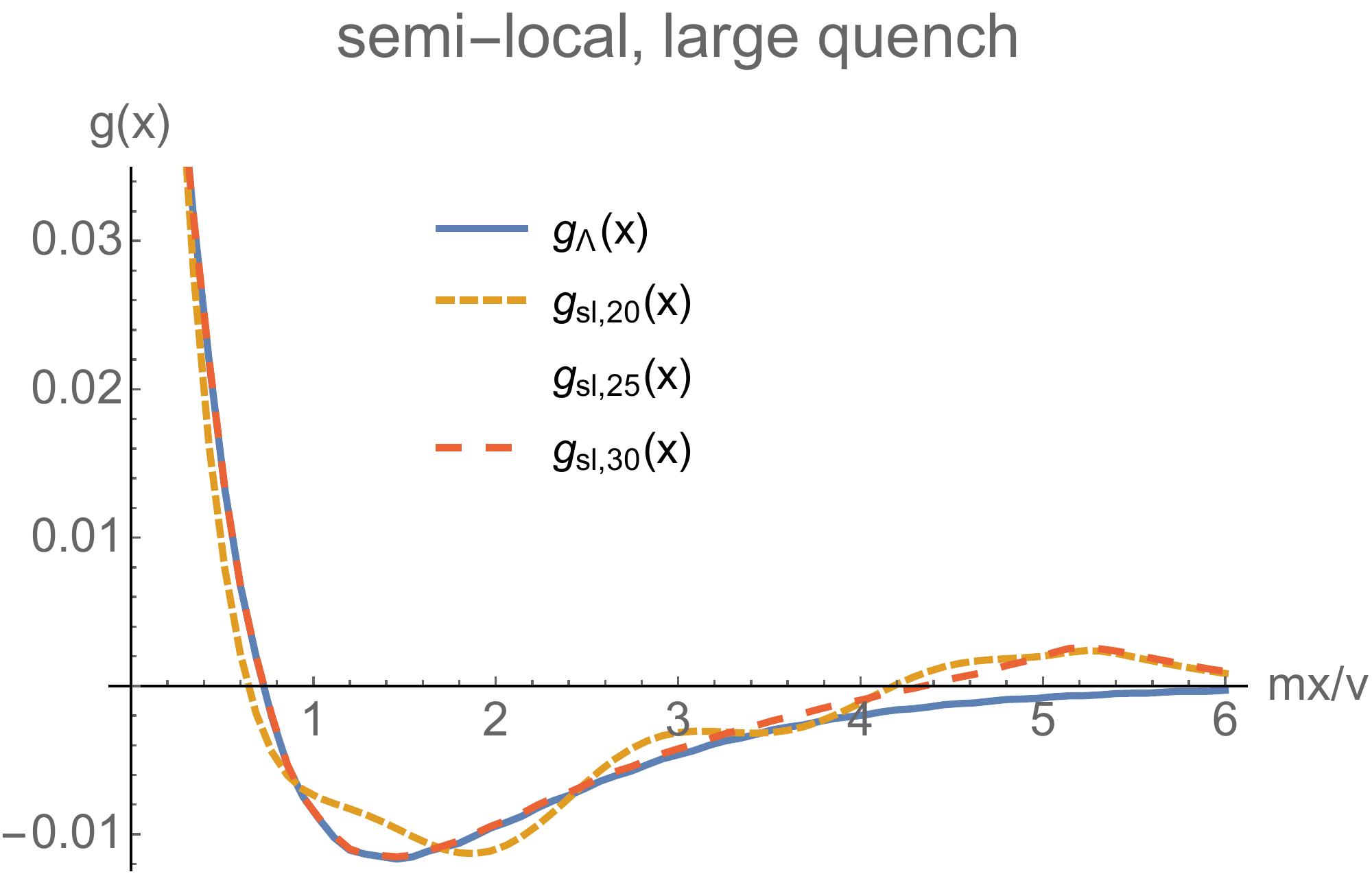}
\end{center}
\caption{$M$-dependence of the truncated semi-local GGE prediction for the
Green's function. The cutoff has been fixed at
$\Lambda=20m/v$ and the spread is $\gamma=5v/m$.}
\label{fig:semi_local_M}
\end{figure}

Our truncated semi-local GGEs depend on three parameters: (i) the number of charges $M$; (ii) the ``spread'' $\gamma$; (iii) the
cutoff~$\Lambda$. We find that, unlike in the ultra-local case, the cutoff dependence does not play an important role and we therefore
exclude it from further discussion. To proceed we simply fix the cutoff $\Lambda$ a sufficiently large value to ensure that the
difference $|g_\Lambda(x)-g(x)|$ is negligible on the interval $[0,x_0]$ of interest. We first consider the $M$-dependence at a fixed
value of $\gamma$, \emph{cf.} Fig.~\ref{fig:semi_local_M}.

\begin{figure}[ht]
  \begin{center}
    \includegraphics[width= 0.65\columnwidth]{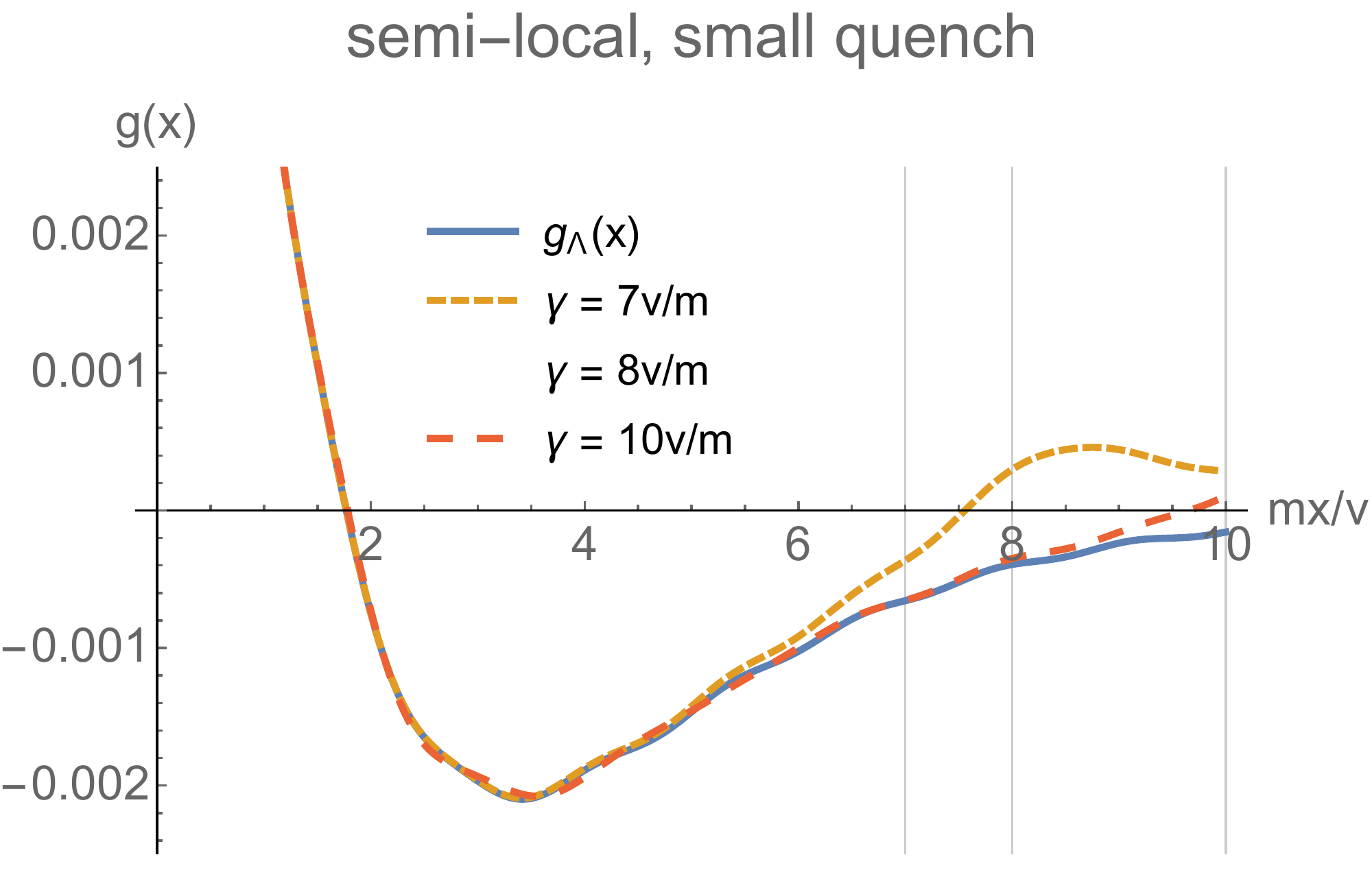}
\end{center}
\caption{$\gamma$-dependence of the truncated semi-local GGE prediction for the
Green's function  $g_{\rm sl,M}(x)$ for fixed number of conserved charges $M=15$. }
\label{fig:semi_local_gamma}
\end{figure}

We observe that, as long as the separation $x$ is sufficiently much smaller than the spread $\gamma$, increasing $M$ leads to a more
accurate description of the stationary state Green's function. As soon as $x\sim\gamma$ increasing $M$ beyond a certain value ceases to
improve the accuracy of the corresponding truncated GGEs. In Fig.~\ref{fig:semi_local_gamma} we show results for the $\gamma$-dependence of the truncated semi-local GGE for a fixed number of charges. We consider a small quench, for which $M=15$ already produces accurate results. We see that the effect of increasing
$\gamma$ is to increase the spatial interval, over which the truncated GGE provides a good approximation to the exact stationary Green's function.

\subsection{Mode occupation number densities}
\begin{figure}[ht]
\begin{center}
  \includegraphics[width= 0.65\columnwidth]{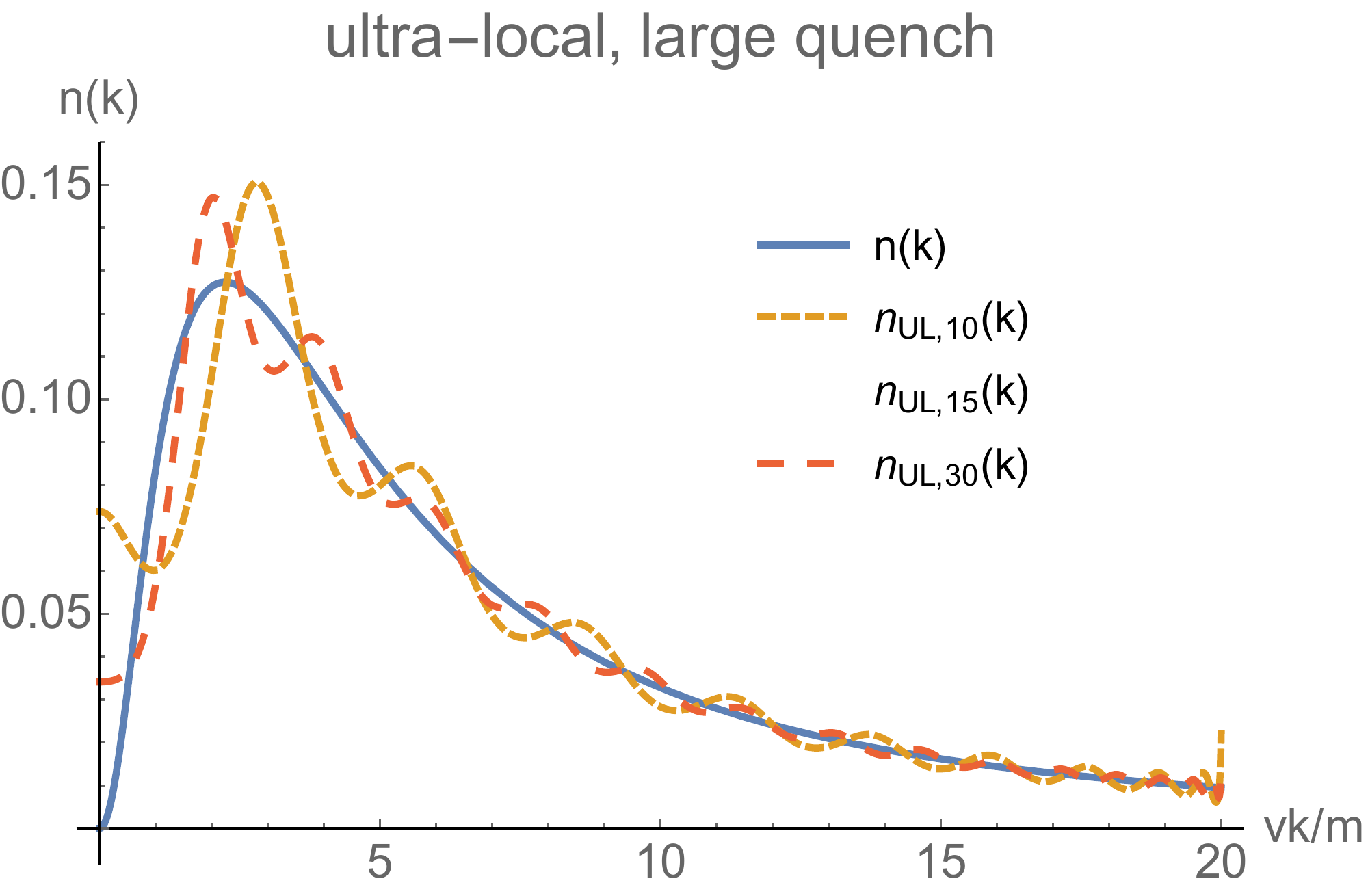}
\end{center}
\caption{
Dependence of the mode occupation densities on the number of charges
in truncated ultra-local GGE.}
\label{fig:momentum_ultra}
\end{figure}
\begin{figure}[ht]
\begin{center}
  \includegraphics[width= 0.65\columnwidth]{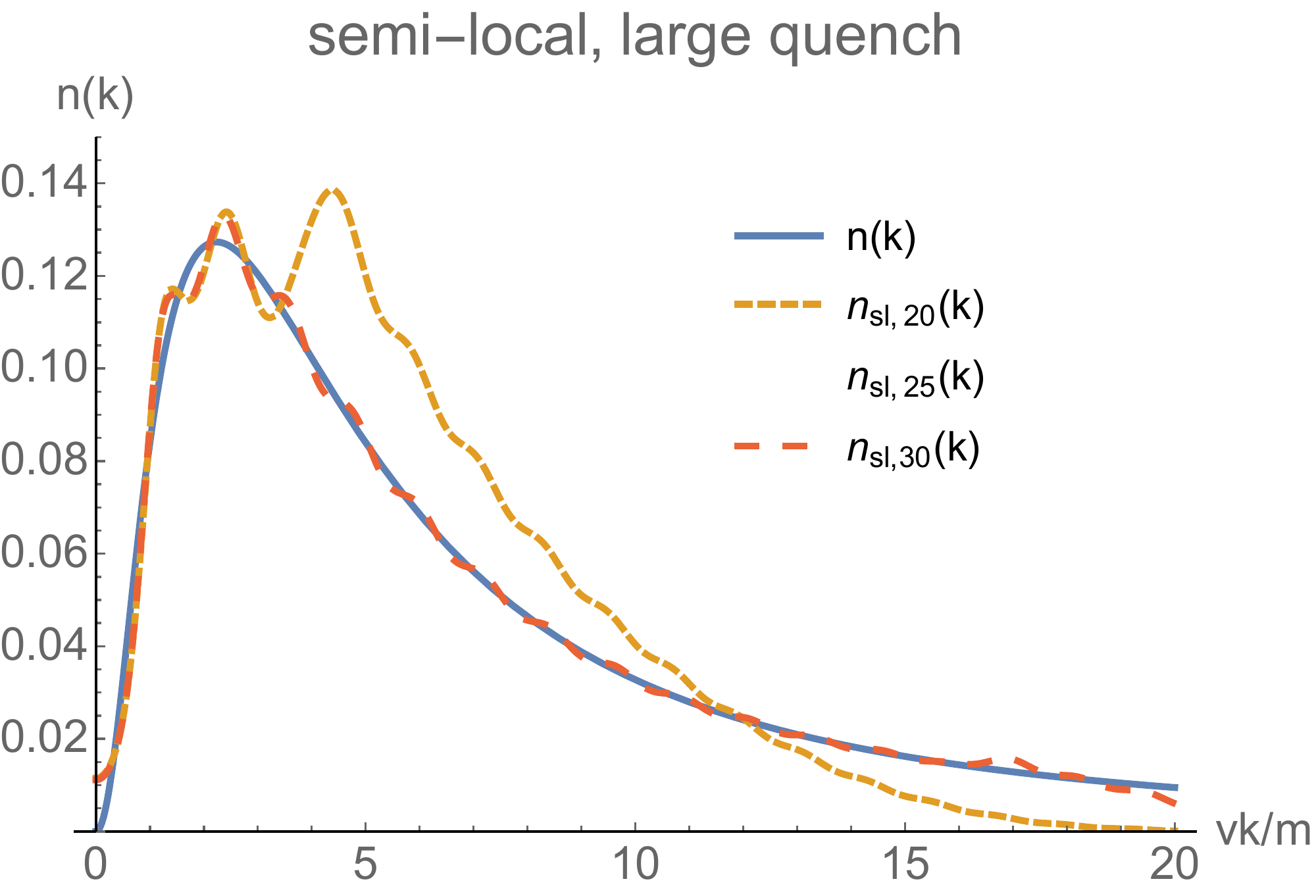}
\end{center}
\caption{
Dependence of the mode occupation densities on the number of charges
in the truncated semi-local GGE with spread $\gamma=5v/m$.}
\label{fig:momentum_semi}
\end{figure}

In the generalized micro-canonical ensemble \cite{GME,QAPRL} the
stationary state after a quantum quench to an integrable model is
described by a simultaneous eigenstate of all conservation laws
\cite{QAPRL}. In general this state is characterized by its Bethe
Ansatz root distribution functions. In our particular case of the
Ising field theory this amounts to specifying the mode occupation
number density $n(k)$. It is therefore an interesting question, how
well our truncated GGEs reproduce the exact $n(k)$ that describes the
stationary state. In Figs.~\ref{fig:momentum_ultra} and~\ref{fig:momentum_semi}
we show results for
large quench. We see that both our truncated GGE's give a
reasonable account of $n(k)$. The quality of the the truncated GGE
predictions for $n(k)$ depends on the number of charges: increasing
the number of ultra-local charges gives a better overall match in the
interval $[0, \Lambda]$ (see Fig.~\ref{fig:momentum_ultra}),
while increasing number of semi-local charges
extends the region of accuracy in $k$ (see Fig.~\ref{fig:momentum_semi}).
We note that above we always considered spreads such that $\gamma >
\Lambda^{-1}$. In this regime the cutoff essentially regularizes the
energy density without influencing other semi-local charges. 
We find that the mode occupation number densities have only a very
weak $\gamma$ dependence. 

\section{Conclusions}
We have considered the implementation of truncated GGEs\cite{FE} for
describing the stationary state after quantum quenches to integrable
field theories, focussing on the simple case of a mass quench in the
Ising field theory. Following Ref.~\cite{pra} we have constructed two
kinds of GGEs, built from ultra-local and semi-local conservation laws
respectively. We have compared the predictions of the two ensembles
for the two-point function $G(x) = \langle \psi^\dagger(x) \psi(0)
\rangle$ of Fermi fields in the stationary state reached at late times
after the quench. We find that once appropriate regularization
procedures have been put in place, both ensembles give a good account
of $G(x)$, although, on balance, the ensemble built from semi-local
charges performs slightly better.
It would be interesting to generalize the results presented here to a
fully interactive field theory such as a Lieb-Liniger
model~\cite{LiLi}. Interaction quenches in this model have been
investigated by a number of groups~\cite{LL}.
A key difficulty compared to the simple case studied here is that
correlations functions of interactive theories as the the Lieb-Liniger
model cannot be calculated analytically, and one needs to resort to
numerical methods such as ABACUS \cite{Abacus}. 

\section*{Acknowledgments}
We thank B. Doyon, D. Fioretto and G. Takacs for helpful
discussions. MP would like to acknowledge hospitality and support from
the Galileo Galilei Institute, where part of this work was carried out
during the  program ``Statistical Mechanics, Integrability and
Combinatorics''. This work was supported by the EPSRC under grant
EP/N01930X/1, the ERC under the Starting Grant no. 279391 and by the
NCN under FUGA grant 2015/16/S/ST2/00448.


\section*{References}


\begin{thebibliography}{99}



\bibitem{RigolPRL07}
M.~Rigol, V.~Dunjko, V.~Yurovsky, and M.~Olshanii, 
Phys. Rev. Lett. {\bf 98}, 050405 (2007).

\bibitem{CalabreseJStatMech07}
P.~Calabrese and  J.~Cardy,  J. Stat. Mech. (2007) P06008.

\bibitem{GGEgeneral}
M.~Cramer, C.~M.~Dawson, J.~Eisert, and T.~J.~Osborne,
Phys. Rev. Lett. {\bf 100}, 030602 (2008);
\nonum T.~Barthel and U.~Schollw\"ock, Phys. Rev. Lett. {\bf 100}, 100601 (2008);
\nonum F.~H.~L.~Essler, S.~Evangelisti, and M.~Fagotti, Phys. Rev. Lett.
{\bf 109}, 247206 (2012);
\nonum B.~Pozsgay, J. Stat. Mech. (2013) {P07003};
\nonum M.~Fagotti and F.~H.~L.~Essler, J. Stat. Mech. (2013) P07012;
\nonum M. Fagotti, J. Stat. Mech. (2014) P03016;
\nonum M.~Fagotti, M.~Collura, F.~H.~L.~Essler, and P.~Calabrese,
Phys. Rev. B {\bf 89}, 125101 (2014);
\nonum B.~Wouters, J.~De~Nardis, M.~Brockmann, D.~Fioretto, M.~Rigol, and
\nonum J.-S.~Caux, Phys. Rev. Lett.  {\bf 113}, 117202 (2014);
\nonum B.~Pozsgay, M.~Mesty\'{a}n, M.~A.~Werner, M.~Kormos, G.~Zar\'{a}nd,
and G.~Tak\'{a}cs, Phys. Rev. Lett.   {\bf 113}, 117203 (2014);
\nonum S.~Sotiriadis and P.~Calabrese, J. Stat. Mech. (2014) {P07024};
\nonum G.~Goldstein and N.~Andrei, Phys. Rev. A {\bf 90}, 043625 (2014);
\nonum M.~Brockmann, B.~Wouters, D.~Fioretto, J.~De.~Nardis, R.~Vlijm, and
J.-S.~Caux, J. Stat. Mech. (2014) P12009;
\nonum B.~Pozsgay, J. Stat. Mech. (2014) P10045;
\nonum M. Rigol, Phys. Rev. E {\bf 90}, 031301(R) (2014);
\nonum M.~Mesty\'an, B.~Pozsgay, G.~Tak\'acs, and M.~A.~Werner,
J. Stat. Mech. (2015) P04001;
\nonum E.~Ilievski, J.~De~Nardis, B.~Wouters, J.-S.~Caux, F.~H.~L.~Essler,
and T.~Prosen, Phys. Rev. Lett. {\bf 115}, 157201 (2015);
\nonum J. Eisert, M. Friesdorf, and C. Gogolin, Nature Phys. 11, 124 (2015);
\nonum E.~Ilievski, E.~Quinn, J.~De~Nardis, and M.~Brockmann,
J. Stat. Mech. (2016) 063101;
\nonum L. Piroli, E. Vernier and P. Calabrese, Phys. Rev. B {\bf 94}, 054313
(2016).

\bibitem{CEF}
P.~Calabrese, F.~H.~L.~Essler, and M.~Fagotti, Phys. Rev. Lett. {\bf
  106}, 227203 (2011);
\nonum
P.~Calabrese, F.~H.~L.~Essler, and M.~Fagotti,
J. Stat. Mech. (2012) P07016;
\nonum
P.~Calabrese, F.~H.~L.~Essler, and
M.~Fagotti, J. Stat. Mech. (2012) P07022.

\bibitem{FE}
M.~Fagotti and F.~H.~L.~Essler, Phys.~Rev.~B {\bf 87}, 245107 (2013).

\bibitem{QAPRL} 
J.-S.~Caux and F.~H.~L.~Essler, Phys. Rev. Lett. {\bf 110}, 257203 (2013).

\bibitem{EF16}
F.~H.~L. Essler and M. Fagotti, J. Stat. Mech. (2016) P064002.


\bibitem{qQFT}
P. Calabrese and J. Cardy, J. Stat. Mech. (2016) 064003;
\nonum
M. A. Cazalilla and M.-C. Chung, J. Stat. Mech. (2016) 064004;
\nonum
P. Calabrese and J. Cardy. Phys. Rev. Lett. {\bf 96}, 136801 (2006);
\nonum
M.~A.~Cazalilla, Phys. Rev. Lett. {\bf 97}, 156403 (2006);
\nonum
V. Gritsev, E. Demler, M. Lukin, and A. Polkovnikov,
Phys. Rev. Lett. 99, 200404 (2007);
\nonum
D.~Fioretto and G.~Mussardo, New J. Phys.  {\bf 12}, 055015 (2010);
\nonum
J.-S.~Caux and R.~M.~Konik, Phys. Rev. Lett. {\bf 109}, 175301 (2012);
\nonum
G.~Mussardo, Phys. Rev. Lett. {\bf 111}, 100401 (2013);
\nonum
M.~Kormos, M.~Collura, and P.~Calabrese, Phys. Rev. A {\bf 89}, 013609 (2014);
\nonum
B. Bertini, D. Schuricht and F.H.L. Essler, J. Stat. Mech. (2014)
P10035;
\nonum
M. Collura, M. Kormos and P. Calabrese, J. Stat. Mech. (2014) P01009;
\nonum
S.~Sotiriadis, Phys. Rev. A {\bf 94}, 031605 (2016);
\nonum
S. Sotiriadis, G. Martelloni, J. Phys. A: Math. Theor. 49 095002 (2016);
\nonum
J.~Cardy, J. Stat. Mech. (2016) 023103;
\nonum
B. Bertini, L. Piroli and P. Calabrese,
J. Stat. Mech. (2016) 063102;
\nonum
T.~Rakovszky, M.~Mesty\'{a}n, M.~Collura, M.~Kormos and G.~Tak\'{a}cs,
arXiv:1607.01068 (2016);
\nonum
A.~Bastianello, S.~Sotiriadis, arXiv:1608.00924 (2016).


\bibitem{LL}
V. Gritsev, T. Rostunov and E. Demler, J. Stat. Mech. (2010) P05012;
\nonum
B. Pozsgay, J. Stat. Mech. (2011) P11017;
\nonum
M.~Panfil, J.~De Nardis and J.-S.~Caux, Phys. Rev. Lett. {\bf 110}, 125302 (2013);
\nonum
M. Collura, S. Sotiriadis and P. Calabrese, Phys. Rev. Lett. {\bf
  110}, 245301 (2013);
\nonum
M. Collura, S. Sotiriadis and P. Calabrese, J. Stat. Mech. (2013) P09025;
\nonum
M. Kormos, A. Shashi, Y.-Z. Chou, J.-S. Caux and A. Imambekov,
Phys. Rev. B {\bf  88}, 205131 (2013);
\nonum
J.~De~Nardis, B.~Wouters, M.~Brockmann, and J.-S.~Caux, 
Phys. Rev. A  {\bf 89}, 033601 (2014);
\nonum
J. De Nardis and J.-S. Caux, J. Stat. Mech. (2014) P12012;
\nonum
D. Iyer and N. Andrei, Phys. Rev. Lett. {\bf 109}, 115304 (2012);
\nonum
J. De Nardis, L. Piroli and J.-S. Caux, Phys. A: Math. Theor. 48 43FT01
(2015);
\nonum
L. Piroli and P. Calabrese, J. Phys. A: Math. Theor. 48, 454002 (2015);
\nonum
L. Piroli, P. Calabrese and F.H.L. Essler, Phys. Rev. Lett. {\bf 116},
070408 (2016);
\nonum
L. Piroli, P. Calabrese and F.H.L. Essler, SciPost Phys. 1, 001 (2016)
[arXiv:1604.08141].


\bibitem{pra}
F.~H.~L.~Essler, G.~Mussardo, and M.~Panfil, 
Phys. Rev. A {\bf 91}, 051602(R) (2015).

\bibitem{Cubero-Vernier} 
E. Vernier and A.C. Cubero, 
arXiv:1609.03220 (2016).


\bibitem{korepin}
V.E. Korepin, A.G. Izergin, and N.M. Bogoliubov, {\em {Quantum Inverse
  Scattering Method, Correlation Functions and Algebraic Bethe Ansatz}}
  (Cambridge University Press, 1993).

\bibitem{Prosen:review}
E.~Ilievski, M.~Medenjak, T.~Prosen and L.~Zadnik, 
J. Stat. Mech. (2016) 064008;
\nonum
E. Ilievski, M. Medenjak and T. Prosen, Phys. Rev. Lett. {\bf 115}, 120601
(2015). 


\bibitem{Doyonthesis}
B. Doyon, \emph{Correlation Functions in Integrable Quantum Field
  Theory}, PhD thesis, Rutgers University (2004).


\bibitem{MajoranaFT}
F. Igl\'oi and H. Rieger, Phys. Rev. Lett. {\bf 85}, 3233 (2000);
\nonum
K. Sengupta, S. Powell, and S. Sachdev, Phys. Rev. A{\bf 69}, 053616 (2004);
\nonum
M. Fagotti and P. Calabrese, Phys. Rev. A {\bf 78}, 010306(R) (2008);
\nonum
D. Rossini, A. Silva, G. Mussardo, G. Santoro, 
Phys. Rev. Lett. {\bf 102}, 127204 (2009);
\nonum
D. Rossini, S. Suzuki, G. Mussardo, G. E. Santoro and A. Silva,
Phys. Rev. B {\bf 82}, 144302 (2010);
\nonum
L. Foini, L.F. Cugliandolo and A. Gambassi, Phys. Rev. B {\bf 84}, 212404
(2011);
\nonum
H. Rieger and F. Igl\'ói, Phys. Rev. B {\bf 84}, 165117 (2011);
\nonum
D. Schuricht and F.H.L. Essler, J. Stat. Mech. P04017 (2012);
\nonum
L. Foini, L.F. Cugliandolo and A. Gambassi, J. Stat. Mech. (2012) P09011;
\nonum
L.~Bucciantini, M.~Kormos, and P.~Calabrese, J. Phys. A:
  Math. Theor. 47 175002.  

\bibitem{GME}
A.~C. Cassidy, C.~W. Clark, and M.~Rigol, Phys. Rev. Lett. {\bf 106},
140405 (2011). 

\bibitem{JS}
J.-S. Caux, J. Stat. Mech. (2016) 064006.



\bibitem{LiLi}
E.H.~Lieb, W.~Liniger, Phys. Rev. {\bf 130}, 1605 (1963).


\bibitem{Abacus} J.S. Caux, J. Math. Phys. 50, 095214 (2009).
\end{thebibliography}
\end{document}